\documentstyle[amsfonts,aps,pre,epsf]{revtex}

\begin{document}
\title{Dynamical entropy for systems with stochastic perturbation}

\author{Andrzej Ostruszka$^1$\footnote{e-mail: ostruszk@if.uj.edu.pl},
Prot Pako{\'n}ski$^1$\footnote{e-mail: pakonski@if.uj.edu.pl},
Wojciech S{\l}omczy{\'n}ski$^2$\footnote{e-mail: slomczyn@if.uj.edu.pl},
and Karol \.Zyczkowski$^{1,3}$\footnote{Fulbright Fellow. Present address:
Centrum Fizyki Teoretycznej PAN, ul. Lotnik\'ow 32/46,
02-668 Warszawa, Poland, e-mail: karol@tatry.if.uj.edu.pl}}

\address{
$^1$Instytut Fizyki im. Mariana Smoluchowskiego, Uniwersytet Jagiello\'nski,
\\ ul. Reymonta 4,  30-059 Krak\'ow, Poland}
\address{
$^2$Instytut Matematyki, Uniwersytet Jagiello\'nski,
\\ ul. Reymonta 4,  30-059 Krak\'ow, Poland}
\address{
$^3$Institute for Plasma Research, University of Maryland,
\\ College Park,  MD 20742, USA}

\date{\today}
\maketitle

\begin{abstract}
  Dynamics of deterministic systems perturbed by random additive noise is
characterized quantitatively. Since for such systems the Kolmogorov-Sinai
(KS) entropy diverges if the diameter of the partition tends to zero, we
analyze the difference between the total entropy of a noisy system and the
entropy of the noise itself. We show that this quantity is finite and
non--negative and call it the dynamical entropy of the noisy system. In the
weak noise limit this quantity is conjectured to tend to the KS-entropy of
the deterministic system. In particular, we consider one-dimensional systems
with noise described by a finite-dimensional kernel for which the
Frobenius-Perron operator can be represented by a finite matrix.
\end{abstract}


\newpage
\section{Introduction}

Stochastic perturbations are typical for any physical realization of a given
dynamical system. Also round-off errors, inevitable in numerical
investigation of any dynamics, may be considered as a random noise.
Quantitative characterization of dynamical systems with external stochastic
noise is a subject of several recent studies
\cite{PSV95,LPPV96,NOA96,LPV96}. On the other hand, the influence of noise
on various low dimensional dynamical systems and the properties of random
dynamical systems have  been extensively studied for many years
\cite{CFH82,De85,MT85,Ka88,BJS90,YOC90}.

Consider a discrete dynamical system generated by $f:X\to X$, where $X$ is a
subset of ${\mathbb R}^d$, in the presence of an additive noise
\begin{equation}
x_{n+1}=f(x_n)+ \xi_n,
\label{langev}
\end{equation}
where $\xi_1, \xi_2, \dots$ are independent random vectors fulfilling
$\langle \xi_n \rangle=0$ and $\langle \xi_n \xi_m \rangle=\sigma^2
\delta_{mn}$. The case with vanishing noise strength $\sigma \to 0$ will be
called the deterministic limit of the model. Properties of such stochastic
systems have recently been analyzed by means of the periodic orbit theory
\cite{CDMV98}. Convergence of invariant measures of the noisy system in the
deterministic limit has been broadly discussed in the mathematical
literature (see for instance
\cite{Ki74,Bo80,Yo86,Ki88,BY93,LM94,Bl97,BK97,BG97}).

A dynamical system generated by $f$ is called chaotic if its
Kolmogorov-Sinai (KS) entropy is positive \cite{CFS82}.  Such a definition
is not applicable for stochastic systems, characterized by infinite entropy.
In this case the partition dependent entropy diverges if the partition
${\cal A}$ of the space $X$ is made increasingly finer.

In this paper we propose a generalization of the KS-entropy for systems with
additive noise (\ref{langev}). Since entropy diverges also for the pure
noise (with the trivial deterministic limit $f(x)=I(x)=x$, for $x \in X$),
we study the difference between the total entropy of the system with noise
and the entropy of the noise itself. Firstly, we set the partition fixed,
and then we take the supremum over all finite partitions with regular cell
boundaries. In this way our definition resembles the {\sl coherent states
dynamical entropy}, two of us proposed several years ago
\cite{SZ94,KSZ97,SZ98} as a measure of quantum chaos. The entropy of the
noise, discussed in this paper, plays the role of entropy of quantum
measurement, connected with the overlap of coherent states and linked to the
Heisenberg uncertainty relation.

Even though our definition is suitable for $d$--dimensional systems with an
arbitrary additive noise, we demonstrate its usefulness on simple
one--dimensional systems. We choose a specific kind of distribution defining
the noise, which can be expanded in a finite basis of $N$ functions in both
variables $x$ and $y$. This condition allows us to express the $n$-steps
probabilities, required to compute the entropy, as a product of certain
matrices. Moreover, we represent the Frobenius--Perron operator of the
system with noise by an $(N+1) \times (N+1)$ matrix, and obtain its spectrum
by numerical diagonalization. The deterministic limit $\sigma \to 0$
requires $N\to \infty$, which resembles the classical limit of quantum
mechanics.

This paper is organized as follows. In Sect.~II the dynamical entropy for
noisy systems is defined and some of their properties are analyzed. One
dimensional systems with expandable noise and their invariant measures are
analyzed in Sect.~III, while different methods of computing the entropy are
presented in Sect.~IV. The entropy for some exemplary systems with noise
(R\'{e}nyi map, logistic map) is studied in Sect.~V. The paper is concluded
by Sect.~VI, while an illustrative iterated function system, used for
computation of the entropy of noise, is provided in Appendix A.

\section{Dynamical entropy for systems with noise}

\subsection{Dynamical entropy for deterministic systems}

Let us consider a partition ${\cal A}=\{E_1,\dots ,E_k\}$ of $X$ into $k$
disjoint cells. The partition generates the symbolic dynamics in the
$k$-symbol code space. Every $n$--steps trajectory can be represented by a
string of $n$ symbols, $\nu=\{i_0,\dots ,i_{n-1}\}$, where each letter $i_j$
denotes one of the $k$ cells. Assuming that initial conditions are taken
from $X$ with the distribution $\mu_f$ invariant with respect to the map
$f$, let us denote by $P_{i_0,\dots ,i_{n-1}}$ the probability that the
trajectory of the system can be encoded by a given string of symbols, i.e.,

\begin{equation}
P_{i_0,\dots ,i_{n-1}} = 
\mu_f \left( \{x : x \in E_{i_0}, f(x) \in E_{i_1}, \dots, 
f^{n-1}(x) \in E_{i_{n-1}} \} \right)
\end{equation}

The {\sl partial entropies}
$H_n$ are given by the sum over all $k^n$ strings of length $n$

\begin{equation}
H_n:= - \sum_{i_0,\dots ,i_{n-1}=1}^k P_{i_0,\dots ,i_{n-1}}
\ln P_{i_0,\dots ,i_{n-1}} ,
\label{part}
\end{equation}
while the {\sl dynamical entropy of the system $f$ with respect to the
partition} ${\cal A}$ reads

\begin{equation}
H(f;{\cal A}):=\lim_{n\to \infty } {\frac 1n} H_n.
\label{padep}
\end{equation}
The above sequence is decreasing and the quantity 

\begin{equation}
H_1 = -\sum_{i=1}^k \mu_f(E_i)\ln[\mu_f(E_i)], 
\label{h1}
\end{equation}
which depends on $f$ only via $\mu_f$, is just the {\sl entropy of the
partition} ${\cal A}$. We denote it by $H_{\cal A}(\mu_f)$. The {\sl
KS--entropy} of the system $f$ is defined by the supremum over all possible
partitions ${\cal A}$ \cite{CFS82}

\begin{equation}
H_{KS}(f):=\sup_{\cal A} H (f;{\cal A}).
\label{parks}
\end{equation}
A partition for which the supremum is achieved is called a {\sl generating}
partition. Knowledge of a $k_g$--elements generating partition for a given
map allows one to represent the time evolution of the system in
$k_g$--letters symbolic dynamics and to find the upper bound for the
KS-entropy: $H_{KS}(f)\le \ln {k_g}$. In the general case it is difficult to
find a generating partition and one usually performs another limit, tending
to zero with the diameter of the largest cell of a partition, which implies
the limit $k\to \infty$. We shall denote this limit by ${\cal A} \downarrow
0$.

\subsection{Entropy for systems with stochastic perturbation}

For simplicity we consider one-dimensional case taking $X=[0,1]$, imposing
periodic boundary conditions and joining the interval into a circle. We
denote the Lebesgue measure on $X$ by $m$, setting $dx=dm(x)$ (clearly
$m(X)=1)$. The noisy system introduced in (\ref{langev}) will be denoted by
$f_{\sigma}$. From now on we assume that all the random vectors $\xi_n$ $(n
\in\mathbb N)$ in (\ref{langev}) have the same distribution with the density
${\cal P}_{\sigma}$. Then the probability density of transition from $x$ to
$y$ under the combined action of the deterministic map $f$ and the noise is
given by ${\cal P}_\sigma(f(x),y) = {\cal P}_\sigma(f(x)-y)$, where $x,y \in
X$ and the difference is taken $\bmod 1$. In the pure noise case ($f=I$)
this density  depends only on the length of the jump and equals to ${\cal
P}_\sigma(x,y) = {\cal P}_\sigma(x-y)$.

We assume that $f_{\sigma}$ has a unique invariant measure
$\mu_{f_{\sigma}}$, which is absolutely continuous with respect to the
Lebesgue measure $m$ (i.e. it has a density $\rho_{f_{\sigma}}$). Clearly,
$\mu_{I_{\sigma}} = m$, and so $\rho_{I_{\sigma}} \equiv 1$. Moreover, we
assume that the measure $\mu_{f_{\sigma}}$ tends weakly to $\mu_f$
respectively, for $\sigma \to 0$, where $\mu_f$ is some invariant measure
for the deterministic system $f$. In Sect.~IIIB we discuss the situation,
where the above assumptions are fulfilled.  

Now, let us fix a partition ${\cal A}$ of $X$. We define the {\sl total
entropy} $H_{tot}(f_{\sigma};{\cal A})$ of the noisy system $f_{\sigma}$ by
formulae (\ref{part}) and (\ref{padep}), analogously to the deterministic
case. Note, however, that in this case the initial conditions should be
taken from $X$ with the measure $\mu_{f_{\sigma}}$. As we shall see below
this entropy grows unboundedly with $k$. Hence, we can not define partition
independent entropy of the noisy system using formula (\ref{parks}), as the
supremum in (\ref{parks}) is equal to the infinity. On the other hand, there
are two kinds of randomness in our model: the first is connected with the
deterministic dynamics; the second comes from the stochastic perturbation.
Accordingly, we split the total partition dependent entropy $H_{tot}$ of a
noisy system $f_{\sigma }$ given by (\ref{part}) and (\ref{padep}) into two
components: the {\sl noise entropy} and the {\sl dynamical entropy}. The
latter quantity characterizes the underlying dynamics $f_{\sigma}$ and is
defined by

\begin{equation}
\label{hdyn}
H_{dyn}(f_{\sigma};{\cal A}) := H_{tot}(f_{\sigma};{\cal A}) -
H_{noise}(\sigma,{\cal A})\, ,
\end{equation}
where the entropy of the noise $H_{noise}(\sigma,{\cal A})$ reads

\begin{equation}
\label{noi-ent}
H_{noise}(\sigma,{\cal A}) = H_{tot}(I_{\sigma};{\cal A}),
\end{equation}
and $I_{\sigma}$ is a stochastic system given by (\ref{langev}) with $f=I$
({\sl pure noise}). Although the both quantities $H_{tot}$ and $H_{noise}$
may diverge in the limit of fine partition ${\cal A} \downarrow 0$
($k\rightarrow\infty$) for a nonzero noise strength, one can make their
difference $H_{dyn}$ bounded, taking an appropriate sequence of partitions,
as we shall see in the next subsection.

In order to keep away from the ambiguity in the choice of a partition, we
eventually define the {\sl dynamical entropy of} $f_{\sigma}$ as

\begin{equation}
H_{dyn}(f_{\sigma}) := \sup_{\cal A}
H_{dyn}(f_{\sigma};{\cal A}),
\label{hdynsup}
\end{equation}
the supremum being taken over all finite partitions ${\cal A}=
\{E_1,\dots,E_k\}$ such that $m(E_i)=1/k$ and $m (\partial E_i) =0$ for
each $i=1,\dots,k$, $k \in \mathbb N$. We will call such partitions {\sl
uniform}. The restriction to uniform partitions is necessary, since
otherwise we may encounter various "pathologies" in the deterministic limit
\cite{O97}. Note, that the uniformity assumption may be omitted in the case
when all the measures $\mu_{f_{\sigma}}$, $\mu_{f}$, and $m$ coincide.

It seems that in many cases the entropy of the noise (\ref{noi-ent}) tends
to zero in the deterministic limit $\sigma\to 0$, and the dynamical entropy
of $f_{\sigma}$ converges to the KS-entropy of the corresponding
deterministic system $f$ (for partial results in this direction see
\cite{Ki88} and \cite{SZ94}, for numerical evidence see Sect.~VD).

\subsection{Boltzmann--Gibbs entropy and bounds for dynamical entropy}

In this section we discuss the behavior of the dynamical entropy in the
another limit, ${\cal A} \downarrow 0$ ($k\rightarrow\infty$). 

Next, we introduce the {\sl Boltzmann--Gibbs (BG) entropy} of the noise:

\begin{equation}
\label{hcgl}
H_{BG}({\sigma}) := -\int_X d\mu_{I_{\sigma}}(x) \int_X dy {\cal P}_\sigma(x-y)
\ln{\cal P}_\sigma(x-y) = -\int_X d\xi {\cal P}_\sigma(\xi)
\ln{\cal P}_\sigma(\xi)\,.
\end{equation}
For interpretation and generalizations of this quantity, sometimes called
{\sl continuous entropy}, consult the monographs of Guia\c{s}u \cite{G77},
Martin and England \cite{ME81}, Jumarie \cite{Ju90}, or Kapur \cite{K94}. In
the simplest case of the rectangular noise given by ${\cal
P}_b(x):=\Theta(x-b/2)\Theta(x+b/2)/b$, for $1 \geq b > 0$ and $x \in X$,
the BG-entropy is equal to ln~$b$. Note, that this quantity vanishes for the
noise uniformly spread over the entire space ($b=1$), becomes negative for
$b<1$ and diverges to minus infinity in the deterministic limit $b\to 0$.

For the system $f_\sigma$ combining the deterministic evolution $f$ and the
stochastic perturbation, the probability density of transition from $x$ to
$y$ during one time step is given by ${\cal P}_\sigma(f(x),y) = {\cal
P}_{\sigma}(f(x)-y)$ for $x,y \in X$. The BG-entropy for this system can be
defined as

\begin{equation}
H_{BG}(f_{\sigma}) = -\int_X d\mu_{f_{\sigma}}(x) \int_X dy
	{\cal P}_\sigma(f(x),y)\ln{\cal P}_\sigma(f(x),y) .
\label{hbol}
\end{equation}
Due to the homogeneity of the noise and due to the periodic boundary
conditions the integral over $y$ in (\ref{hbol}) does not depend on $x$.
Therefore for any system $f$ perturbed by a nonzero noise ($\sigma > 0$) one
obtains

\begin{equation}
\label{hceqhcf}
H_{BG}(f_\sigma) \equiv H_{BG}(\sigma) =
-\int_X dy {\cal P}_\sigma(y) \ln{\cal P}_\sigma(y)  \,.
\end{equation}
Applying this equality and using the same methods as in \cite{SZ98} we can
prove that the total entropy fulfills the following inequalities (the first
inequality can be deduced from the lower bound for the variation of
information obtained in Theorem~2.3 from \cite{G77}; the second inequality
comes from the definition) 

\begin{equation}
\label{bgineq}
H_{BG}(\sigma) + \ln k \leq
H_{tot}(f_\sigma,{\cal A}) \leq
H_{\cal{A}}(\mu_{f_{\sigma}}) \,.
\end{equation}
For $f=I$ we get

\begin{equation}
\label{bgineqnoise}
H_{BG}(\sigma) + \ln k \leq
H_{noise}(\sigma,{\cal A}) \leq
\ln k \,.
\end{equation}

Hence, both the total entropy $H_{tot}(f_\sigma,{\cal A})$ and the noise
entropy $H_{noise}(\sigma,{\cal A})$ diverges logarithmically in the limit
${\cal A} \downarrow 0$. Let us now study how does the dynamical entropy,
which is the difference of these quantities, depend on the partition ${\cal
A}$.

If the partition ${\cal A}=\{X\}$ consists of one cell only ($k=1$), we have
$H_{tot}(f_\sigma,{\cal A}) = H_{noise}(\sigma,{\cal A}) = H_{\cal A} = 0$.
Thus, the dynamical entropy with respect to this trivial partition equals
zero for any system, which guarantees that the dynamical entropy  given by
the supremum over all partitions (\ref{hdynsup}) is non--negative.

Let us now investigate the behavior of the total entropy in the opposite
case ${\cal A} \downarrow 0$ ($k\rightarrow\infty$) for a non-zero noise
strength $\sigma$.  Performing the time limit (\ref{padep}) we find, as in
\cite{SZ98}, that for very fine partitions the total entropy of the system
is given approximately by the sum of the Boltzmann-Gibbs entropy and the
entropy of the partition (this statement is again based on the Theorem~2.3
from \cite{G77})

\begin{equation}
\label{totapprox}
H_{tot}(f_\sigma,{\cal A}) \stackrel{{\cal A}\downarrow 0}{\approx}
H_{BG}(\sigma) + \ln k \,.
\end{equation}
Observe that due to the property (\ref{hceqhcf}) the right-hand side does
not depend on the dynamical system $f$ and the approximate equality
(\ref{totapprox}) holds also for the entropy of the noise
$H_{noise}(\sigma,{\cal A})$. Therefore dynamical entropy tends to zero for
both limiting cases

\begin{eqnarray}
H_{dyn}(f_\sigma,\{X\}) &=& 0 ~~~~~~ ({\rm for} ~~ k=1)\\
\lim_{{\cal A}\downarrow 0}H_{dyn}(f_\sigma,{\cal A}) &=&
0 ~~~~~~ ({\rm for} ~~ k \to \infty) \label{limdyn} \,.
\end{eqnarray}

Let us now discuss, what are the minimal and maximal dynamical entropies
admissible for a certain kind of stochastic noise. From (\ref{bgineq}) and
(\ref{bgineqnoise}) we get

\begin{equation}
\label{bgineq2}
H_{BG}(\sigma) \leq
H_{dyn}(f_\sigma,{\cal A}) \leq
- H_{BG}(\sigma) - \ln k + H_{\cal A}(\mu_{f_{\sigma}}) \leq
- H_{BG}(\sigma) \,.
\end{equation}
Thus the dynamical entropy is bounded from above by $-H_{BG}(\sigma)$.
Combining this with (\ref{limdyn}) one obtain

\begin{equation}
0 \leq H_{dyn}(f_\sigma) \leq -H_{BG}(\sigma)
\label{hmax}
\end{equation}
This relation provides a valuable interpretation of the Boltzmann--Gibbs
entropy. This quantity, determined by the given probability distribution of
the noise ${\cal P}_\sigma$, tells us whether the character of the dynamics
of a specific deterministic system $f$ can be resolved under the influence
of this noise. For example, the rectangular noise of width $b=1$ may be
called {\sl disruptive}, since the corresponding  BG--entropy is equal to
zero, and consequently $H_{dyn}(f_\sigma, {\cal A}) = 0$ for every uniform
partition ${\cal A}$. Under the influence of such a noise we have no
information, whatsoever, concerning the underlying dynamics $f$.
Furthermore, it is unlikely to distinguish between two systems, both having
KS--entropies larger than $-H_{BG}(\sigma)$ of the noise present.

Evidently, in the deterministic limit the maximal entropy tends to infinity.
On the other hand, in this case, one obtains in (\ref{limdyn}) the
KS--entropy. This apparent paradox consists in the order of the two limits:
the number of cells in coarse-graining to infinity and the noise strength to
zero. These two limits do not commute.

Note that several authors proposed different approaches to the notion of
dynamical entropy of noisy system. Crutchfield and Packard \cite{CP83}
introduced the {\sl excess entropy} to analyze the difference between
partial entropies of a noisy system and the corresponding deterministic
system, and investigated its dependence on the noise strength $\sigma$ and
the number of the time steps $n$.

In order to avoid problems with the unbounded growth of the total entropy
for sufficiently fine partitions Gaspard and Wang studied
$\epsilon$--entropy \cite{GW93}, where the supremum is taken only over the
class of the partitions, for which the minimal diameter of a cell is larger
than $\epsilon$. This quantity can be numerically approximated by the
algorithm of Cohen and Procaccia \cite{CP85}. The $\epsilon$-entropy
diverges logarithmically in the limit $\epsilon \to 0$; the character of
this divergence may be used to classify various kinds of random processes
\cite{GW93,Ga98}.

The dependence of the dynamical entropy on time yields another interesting
problem. For discrete deterministic systems the KS-entropy is additive in
time: $H_{KS}(f^T)=T H_{KS}(f)$. On the other hand it follows from
(\ref{hmax}) that the dynamical entropy of a noisy system fulfills
$H_{dyn}(f_{\sigma}^T) \leq -H_{BG}(\sigma)$, for each time $T$. Thus for a
nonzero $\sigma$ the ratio $H_{dyn}(f_{\sigma}^T)/T$ tends to zero in the
limit $T\to \infty$, while for the deterministic dynamics
$[H_{KS}(f^T)]/T=H_{KS}(f)$. The symbol $f_{\sigma}^T$ represents the same
deterministic system $f$, subjected to the stochastic perturbation only once
for $T$ time steps. The related issue has been recently raised by Fox
\cite{Fo95} in the context of deterministic evolution of Gaussian densities.
The discontinuity of $H_{dyn}(f_{\sigma}^T)/T$ in the limit $\sigma \to 0$
is a consequence of the fact that the another limits: time to infinity and
noise strength to zero do not commute.

In some sense this resembles the noncommutativity of the limits {\sl{time}}
$\to \infty$ and $\hbar \to 0$ in quantum mechanics, crucial for
investigation of the so--called {\sl quantum chaos} (see e.g.
\cite{CC95,ASAA96}). Continuing this analogy even further, the entropy of
noise corresponds to the entropy of quantum measurement \cite{SZ94,KSZ97},
while the Boltzmann-Gibbs entropy $H_{BG}$ plays the role of the Wehrl
entropy \cite{We79}, recently used by two of us (WS, K\.Z) to estimate the
coherent states dynamical entropy \cite{SZ98}.

\subsection{Systems with rectangular noise}

We now discuss the computation of the entropy of noise for the rectangular
noise ${\cal P}_b$ (see Sect.~IIC), with the periodic boundary conditions
imposed. Computation of the transition probabilities in (\ref{part}) reduces
to simple convolutions of the rectangular noise and is straightforward for
the first few time steps. For larger $n$ the calculations become tedious,
and the convergence in the definition of entropy (\ref{padep}) is rather
slow (not faster than $1/n$). It is hence advantageous to consider the
sequence of {\sl relative entropies} which converge much faster to the same
limit $H(f;{\cal A})$ \cite{CP83,CP85}. For some systems the exponential
convergence of this quantity was reported \cite{SG86,MZ87,ZS95}.

In our analytical and numerical computations we used relative entropies
$G_n$. In all of the cases studied, the term $G_7$ gives the entropy
$\lim_{n\to \infty} G_n$ with a relative error smaller than $10^{-5}$. For a
partition consisting of two equal cells ($k=2$) and the rectangular noise
$P_b$ we obtained an explicit expression for $G_4$, as a function of the
noise width $b$. Analytical result obtained in \cite{O97} are too lengthy to
reproduce here gives an approximation of the entropy of noise with precision
$10^{-4}$.

We analyzed the dynamical entropy of the R\'{e}nyi map $f_s(x)=[sx]_{\bmod
1}$ (with integer parameter $s$) subjected to the rectangular noise.
Independently of the noise strength, the uniform distribution remains the
invariant density of this system. For a large noise $b\sim 1$ the dynamical
entropy is close to zero, since the difference between the noise and the
system with noise is hardly perceptible. The dynamical entropy grows with
the decreasing noise width $b$ and in the deterministic case seems to tend
to the KS-entropy of the R\'{e}nyi map $H_{KS}(f_s)=\ln s$. For more
involved systems the computation of the dynamical entropy becomes rather
difficult even for this simple rectangular noise. In order to avoid
calculating $k^n$ different integrals in (\ref{part}), in the subsequent
section we introduce the class of noises for which computing of
probabilities $P_{i_0,\dots,i_{n-1}}$ for any dynamical system reduces to
multiplication of matrices.

\section{Systems with smooth noise of discrete strengths}

\subsection{Model distribution of noise}

In this section we define the particular discrete family of the probability
distributions $P_N$ representing the noise and study properties of dynamical
system subjected to this noise. As above, we consider one-dimensional space
$X=[0,1)$ and impose periodic boundary conditions. We shall look for a
kernel ${\cal P}(x,y)$ homogeneous, periodic, and being decomposable in a
finite basis
\begin{eqnarray}
{\cal P}(x,y) & \equiv & {\cal P}(x-y)={\cal P}(\xi),  \nonumber \\
{\cal P}(x,y) & \equiv & {\cal P}(x\bmod1,y\bmod1), \nonumber \\
{\cal P}(x,y) & = &  \sum_{l,r=0}^{N}A_{lr}u_r(x)v_l(y),
\label{expand}
\end{eqnarray}
for $x,y \in \mathbb R$, where $A = (A_{lr})_{l,r = 0, \dots ,N}$ is a real
matrix of expansion coefficients. We assume that the functions $u_r; ~r =
0,\dots,N$ and $v_l; ~l = 0,\dots,N $ are continuous in $X=[0,1)$ and
linearly independent. Consequently, we can uniquely express $f \equiv 1$ as
their linear combinations. Both sets of base functions form an
$(N+1)$-dimensional Hilbert space. The last property in (\ref{expand}) is
necessary in order to proceed with the matrix method of computation of the
probabilities in (\ref{part}).

All these conditions are satisfied by the {\sl trigonometric noise}
\begin{equation}
\label{trignoise}
{\cal P}_N(\xi) = {\rm C_N}\cos^N(\pi\xi)\,,
\end{equation}
where $N$ is even ($N=0,2,\dots$).  The normalization constant ${\rm C_N}$
can be expressed in terms of the Euler beta function $B(a,b)$ or the double
factorial
\begin{equation}
{\rm C_N} = \frac{\pi}{B(\frac{N+1}{2},\frac{1}{2})}\ = \frac{N!!}{(N-1)!!}.
\label{cnorm}
\end{equation}
We use basis functions given by
\begin{eqnarray}
u_r(x) &=& \cos^r(\pi x) \sin^{N-r}(\pi x), \nonumber\\
v_l(x) &=& \cos^l(\pi y) \sin^{N-l}(\pi y),
\end{eqnarray}
where $x \in X$ and $r,l = 0,\dots,N$. We do not require their
orthonormality. Expanding cosine as a sum to the $N$-th power in
(\ref{trignoise}) we find that the $(N+1) \times (N+1)$ matrix $A$, defined
in (\ref{expand}), is diagonal for this noise
\begin{equation}
A_{lr}= {\rm C_N} {N \choose l} \delta_{lr}.
\label{alr}
\end{equation}

The parameter $N$ controls the strength of the noise measured by its
variance
\begin{equation}
\sigma^2 = \frac{1}{2\pi^2}\Psi'(\frac{N}{2}+1)
 = \frac{1}{2\pi ^{2}}\left( \sum\limits_{k=\left(N/2\right)
  +1}^{\infty }\frac{1}{k^{2}}\right),
\end{equation}
where $\Psi'$ stands for the derivative of the digamma function \cite{GR65}.
%FIG 1
\begin{figure}
\centering
\setlength{\epsfxsize}{0.6\textwidth}
\epsfbox{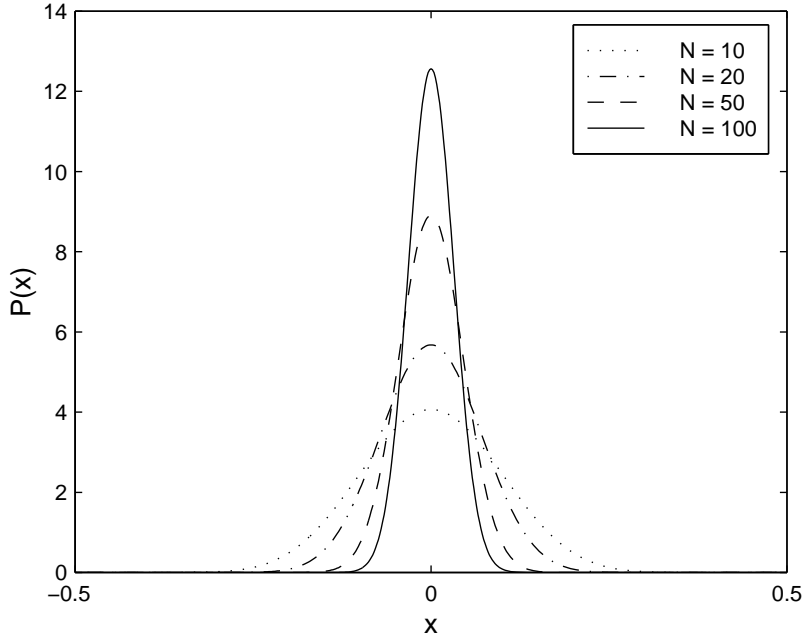}
\caption{Probability density of the noise ${\cal P}(x)$ for $N=10,20,50$ and
$100$.}
\label{figtrig}
\end{figure}
%FIG 2
\begin{figure}
\centering
\setlength{\epsfxsize}{0.6\textwidth}
\epsfbox{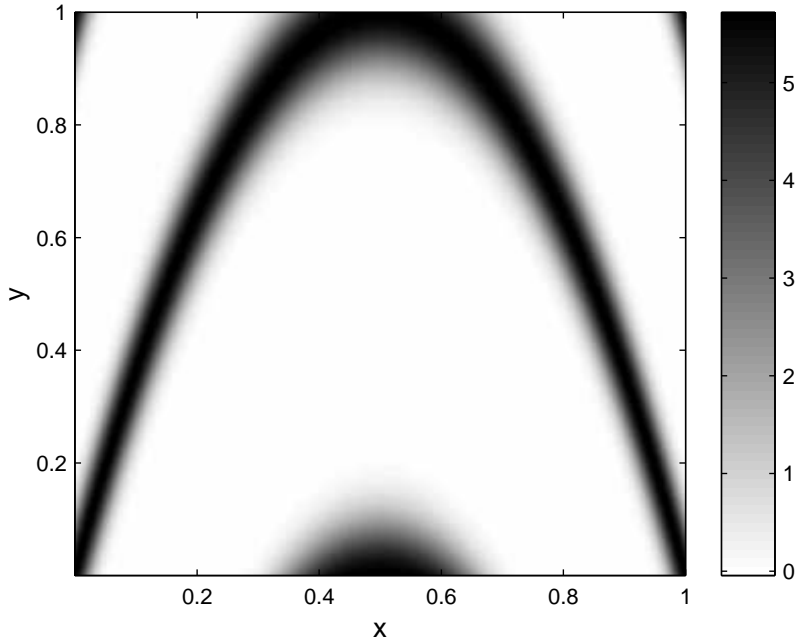}
\caption{Transition kernel ${\cal P}_N(f(x),y)$ for the logistic map
$f(x)=4x(1-x)$ with the noise characterized by $N=20$. Darker colour denotes
higher value of the kernel according to the attached scale. The variable $x$
is periodic; $x=x_{\bmod 1}$.}
\label{logdens}
\end{figure}
Fig.~\ref{figtrig} presents the densities of the noise for $N=10,20,50$ and
$100$. The deterministic limit is obtained by letting $N$ tend to infinity.
Since $N$ determines the size of the Hilbert space, in which the evolution
of the densities takes place, it can be compared with the quantum number
$j\sim 1/\hbar$ used in quantum mechanics. Note that for every value of the
parameter $N$ the probability function ${\cal P}_N(x)>0\,~\mbox{for}~x\neq
1/2$ (mod $1$), so the analyzed perturbation is not {\sl local} in the sense
of Blank \cite{Bl97}.

It is worthwhile to mention that the properties (\ref{expand}) are preserved
for the kernel ${\cal P}_N(f(x),y)$ describing the dynamics of the system
with noise (\ref{langev}). The expansion matrix $A$ is the same, if one use
the modified basis functions defined by $\widetilde{u}_k(x):=u_k(f(x))$, for
$x \in X$, which explicitly depend on the deterministic dynamics $f$. To
illustrate some features of our model we plot in Fig.~\ref{logdens} the
transition kernel ${\cal P}_N(f(x),y)$ for the logistic map perturbed by the
noise defined in (\ref{trignoise}) with $N=20$.
%FIG 3
\begin{figure}
\centering
\setlength{\epsfxsize}{0.6\textwidth}
\epsfbox{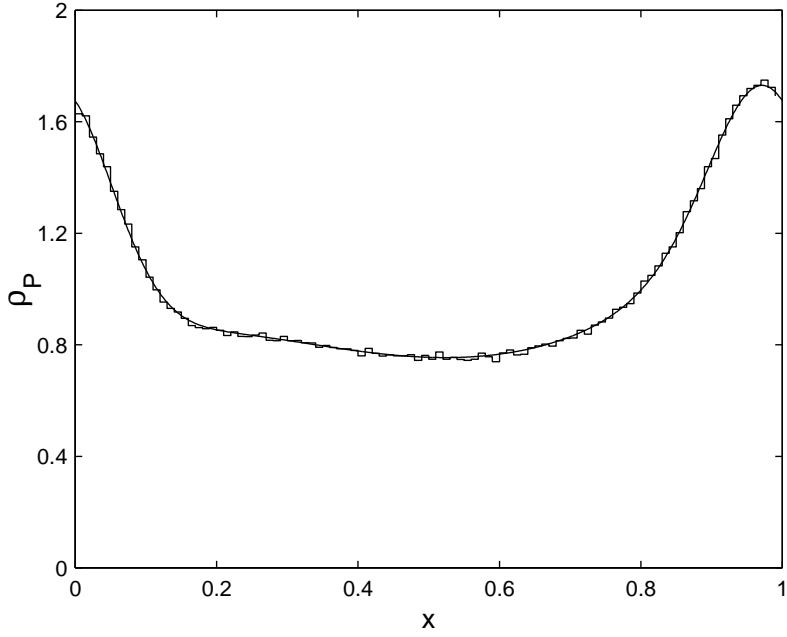}
\caption{Invariant density for the logistic map $f(x)=4x(1-x)$ subjected to
the trigonometric noise ($N=20$): solid line represents the leading
eigenvector of the matrix $D$; histogram is obtained by iteration of one
million of initial points by the noisy map.}
\label{rolog}
\end{figure}
%FIG 4
\begin{figure}
\centering
\setlength{\epsfxsize}{0.6\textwidth}
\epsfbox{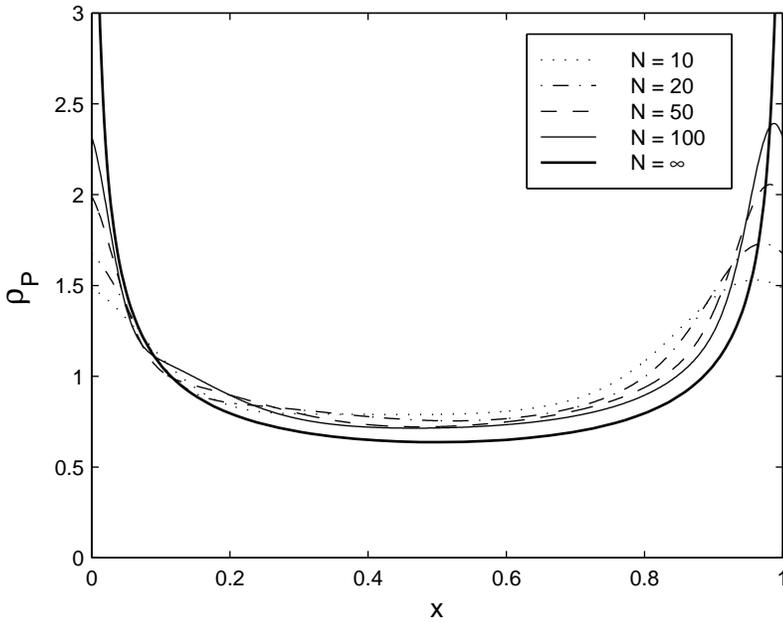}
\caption{Invariant densities of the logistic map for parameters of the
trigonometric noise $N=10,20,50$ and $100$, together with the deterministic
limit $N\rightarrow\infty$.}
\label{roforlog}
\end{figure}

\subsection{Invariant measure for systems with stochastic perturbation}
\label{invmeasure}

The density of the invariant measure $\rho_P$ of the system (\ref{langev})
is given as the eigenstate of the Frobenius--Perron operator $M_P$
corresponding to the largest eigenvalue equal $1$. For the deterministic
system, the invariant density $\rho$ fulfills the formal equation

\begin{equation}
\rho(y)=\int_0^1 \delta(f(x)-y)\rho(x)dx\,.
\label{eig1}
\end{equation}
In the presence of a stochastic perturbation this equation becomes

\begin{equation}
\label{eigenro}
\rho_P(y) = (M_P(\rho_P))(y) = 
\int_0^1 \int_0^1 {\cal P}(x',y) \delta (f(x)-x') \rho_P(x) dx' dx =
\int_0^1 {\cal P}(f(x),y)\rho_P(x)dx\,,
\end{equation}
where $M_P$ is the {\sl Frobenius-Perron (FP) operator} connected with the
noisy system (\ref{langev}). Let us assume that the kernel satisfies the
conditions (\ref{expand}) listed in the preceding subsection and it can be
expanded as ${\cal P}(f(x),y)=\sum_{l,r=0}^N A_{lr} u_r(f(x)) v_l(y)$. Then
we have

\begin{equation}
\begin{array}{rcl}
M_P(\rho)(y)& = & \displaystyle\int_0^1
	\sum_{l,r=0}^N A_{lr} u_r(f(x)) v_l(y) \rho(x) dx
  = \sum_{l,r=0}^N A_{lr} \bigl[\int_0^1 u_r(f(x)) \rho(x) dx \bigr] v_l(y) \\
& = & \displaystyle\sum_{r=0}^N
	\bigl[\int_0^1 u_r(f(x)) \rho(x) dx \bigr] \tilde{v}_{r}(y)
\end{array}
\label{fp1}
\end{equation}
for $y \in X$, where $\tilde{v}_r=\sum_{l=0}^NA_{lr}v_l$. Thus, any initial
density is projected by the FP--operator $M_P$ into the vector space spanned
by the functions $\tilde{v}_{r}; r=0,\dots,N$, and so its image may be
expanded in the basis $\{\tilde{v}_{l}\}_{l=0,\dots,N}$. This statement
concerns also the invariant density $\rho_P$. Expanding $\rho_P$
\begin{equation}
\rho_P = \sum_{l=0}^N {q(P)}_l \tilde{v}_l
\label{invmes}
\end{equation}
with unknown coefficients  ${q(P)}_l$ and inserting this into (\ref{fp1}) we
obtain the eigenequation for the vector of the coefficients
${q(P)}=\{{q(P)}_0,\dots,{q(P)}_N\}$
\begin{equation}
{q(P)} = D{q(P)}.
\label{fp2}
\end{equation}
The FP-operator is represented by the $(N+1)$ dimensional matrix $D=BA$,
where $A$ is given by (\ref{alr}), and the entries of the matrix $B$ are
given by
\begin{equation}
B_{rm} = \int_0^1 u_r(f(x)) v_m(x) dx.
\label{brm}
\end{equation}
for $n,m=0,\dots,N$. Observe that $A$ does not depend on the deterministic
dynamics $f$, while $B$ depends on the noise via the basis functions $u$ and
$v$. Furthermore, note that
\begin{equation}
D_{rm} = \int_0^1 u_r(f(x)) \tilde{v}_m(x) dx
\label{drm}
\end{equation}
for $r,m=0,\dots,N$, and
\begin{equation}
M_P(\sum_{l=0}^N {q(P)}_l \tilde{v}_l(y)) =
	\sum_{l=0}^N (D{q(P)})_l\tilde{v}_l(y)
\label{fp3}
\end{equation}
for each vector ${q(P)}=\{{q(P)}_0,\dots,{q(P)}_N\} \in {\mathbb R}^{N+1}$.
It follows from (\ref{fp1}) and (\ref{fp3}) that there is a one-to-one
correspondence between the eigenvectors of $D$ and the eigenfunctions of the
FP--operator $M_P$. The latter has a one--dimensional eigenspace
corresponding to the eigenvalue $1$, since the kernel ${\cal P}(x,y)$
vanishes only for $x-y=1/2$ (mod $1$), which implies that the two-step
probability $\int{\cal P}(x,z){\cal P}(z,y)dz>0$ for $x,y \in X$ (see
\cite{LM94},~Th.~5.7.4). Thus the equation (\ref{fp2}) has the unique
solution $q(P)$ fulfilling
\begin{equation}
\sum_{r=0}^N {q(P)}_r \int_0^1 \tilde{v}_r(y)dy = 1 \,,
\label{norm}
\end{equation}
or equivalently $\langle q(P), \tau \rangle = 1$, where $\tau=
\bigl(\int_X\tilde{v}_0(y)dy, \int_X\tilde{v}_1(y)dy, \ldots,
\int_X\tilde{v}_N(y)dy \bigr)$. We find it diagonalizing numerically the
matrix $D$. The function $\rho_P$ given by (\ref{invmes}) is then the
invariant density for the system with noise $f_{\sigma}$.

This technique was used to find the invariant measure for the logistic map
given by $f(x)=4x(1-x)$, for $x \in X$ in the presence of noise.
Fig.~\ref{rolog} presents the invariant density for the logistic map with
noise parameter $N=20$. It can be compared to the histogram showing the
density of the $11$-th iteration of one million uniformly distributed random
initial points.

In the deterministic limit $\sigma \to 0$ the size of the matrix $N+1$ grows
to infinity. We believe that our approach can be used to approximate the
invariant measure of the deterministic system by decreasing the noise
strength. Fig.~\ref{roforlog} presents a plot of invariant densities for the
logistic map perturbed with the trigonometric noise for $N=10,20,50$ and
$100$, compared with the invariant measure for the deterministic case
$N\to\infty$ given by \cite{O93} \begin{equation}
\rho(y)=\frac{1}{\pi\sqrt{y(1-y)}}\, \end{equation} for $y \in X$.

\subsection{Spectrum of randomly perturbed systems}

The spectral decomposition of the Frobenuis--Perron operators corresponding
to classical maps is a subject of an intensive current research
\cite{Ga92,AT92,HS92,Ga93,HD94,LM94,Fo95,Fo97}. The spectrum of a
FP-operator is contained in the unit disk on the complex plane and depends
on the choice of a function space, in which acts the FP-operator. If the
dynamical system has an invariant density exists, the largest eigenvalue is
equal to the unity. The radius of the second largest eigenvalue determines
the rate of convergence to the invariant measure. To characterize the
spectrum one defines {\sl essential spectral radius} $r$. It is the smallest
non-negative number, for which the elements of the spectrum outside the disk
of radius $r$, centered at the origin, are isolated eigenvalues of finite
multiplicity. It was shown \cite{Ke84} that for one-dimensional piecewise
$C^2$ expanding maps and the FP-operator defined on the space of functions
of bounded variations, the spectral radius is related to the expanding
constant.

\begin{figure}
\centering
\setlength{\epsfxsize}{0.5\textwidth}
\epsfbox{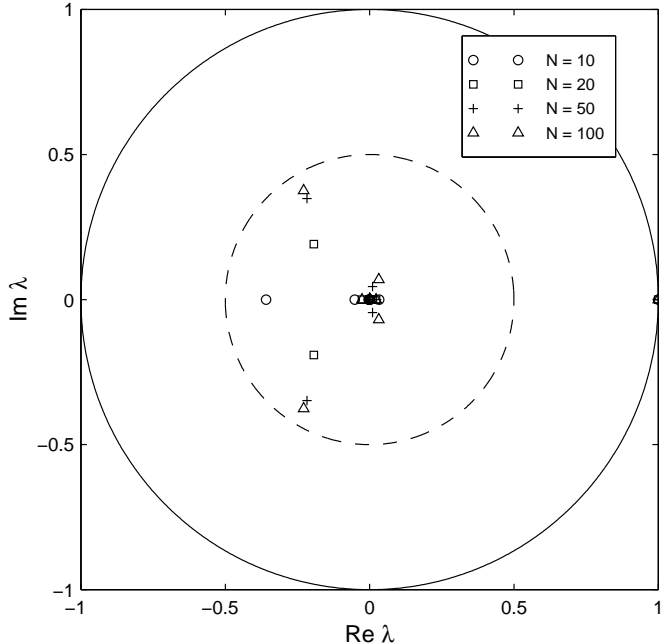}
\caption{Spectra of the FP-operator for the logistic map $f(x)=4x(1-x)$
subjected to the noise ${\cal P}_N$ with $N= 10,20,50$ and $100$.}
\label{spect}
\end{figure}

We analyzed the spectral properties of the FP-operator of the perturbation
of the logistic map, for which the Lyapunov exponents equals to ln(2) and
$r=1/2$. The interval $[0,1]$ is joined into a circle to keep the system
conservative in the presence of noise. FP-operator of the system subjected
to the shift-invariant additive perturbation ${\cal P}_N$ is represented by
the matrix $D$ of the size $N+1$. We obtained its spectrum by the numerical
diagonalization. The largest eigenvalue $\lambda_1$ of $D$ was equal to the
unity up to a numerical error of order $10^{-10}$. The second eigenvalue
$\lambda_2$ was found to approach the essential spectral radius $r$ in the
deterministic limit $N\to\infty$. Since the matrix $D$ has real entries, its
eigenvalues are real or appear in conjugate pairs. Fig.~\ref{spect} presents
the largest eigenvalues of this system for $N=10,20,50$ and $100$. All other
eigenvalues are so small that they coincide with the origin in the picture.
Observe, that eigenvalues do not tend to the values $\lambda_m=1/4^{m-1}$
for $m=1,2,\dots$ found for the deterministic system in \cite{Fo95}.

Our results show that the structure of the spectrum of the FP-operator of a
deterministic system depends on the character of the method used to
approximate it. Introducing a random noise may be considered as a possible
approximation, since it enables us to represent the FP-operator by a matrix
of a finite dimension. In other words, the presence of the noise
predetermines a certain space, in which the eigenstates live. This numerical
finding corresponds to the recent results of Blank and Keller \cite{BK98},
who showed the instability of the spectrum for some maps subjected to
certain perturbations.

%FIG 5
\section{Computing entropy for systems with expandable noise}

\subsection{Matrix formulation of probability integrals}

Our aim is to compute probabilities entering the definition of the total
dynamical entropy of a noisy system $P_{i_0,\dots,i_{n-1}}$
\begin{equation}
\label{probab}
P_{i_0,\dots,i_{n-1}} = \int_{E_{i_0}}\!\!\!\!\!\rho_P(x_0)\, dx_0
\int_{E_{i_{1}}}\!\!\!\!\! dx_1 \ldots
\int_{E_{i_{n-1}}}\!\!\!\!\!\!\!\! dx_{n-1}
{\cal P}(f(x_0),x_1){\cal P}(f(x_1),x_2)\cdots{\cal P}(f(x_{n-2}),x_{n-1})\,.
\end{equation}
Introducing $(n-1)$ times the expansion (\ref{expand}) applied to the kernel
${\cal P}(f(x),y)$ and interchanging the order of summing and integration we
arrive at
\begin{equation}
P_{i_0,\dots,i_{n-1}} =
\tau^T \bigl[ D(i_{n-1}) \cdots D(i_1)D(i_0) \bigr] q(P)\,,
\label{mat1}
\end{equation}
where $A$, $\tau$ and $q(P)$ are defined in Sect.~IIIB,  $D(i)=B(i)A$, and
matrices $B(i)$ are given by the integral over the cell $E_i$, i.e.
$B(i)_{rl}=\int_{E_i}\!\!u_r(f(x))v_l(x)dx$ for  $i=1,\dots,k$;
$r,l=0,\dots,N$ in the analogy to (\ref{brm}).

The above formula provides a significant simplification in the computation
of entropy. Instead of performing multidimensional integrals in
(\ref{probab}), we start from computing the matrices $D(i)$ for any cell
$i=1,\dots,k$, and receive the desired probabilities by matrix
multiplications. By this method the probabilities may be efficiently
obtained even for larger numbers of the time steps $n$. The only problem
consists in the number of terms in (\ref{part}), equal to $k^n$, which for
larger number of cells $k$ becomes prohibitively high. To overcome this
difficulty we apply in this case the technique of iterated functions systems
presented below.

\subsection{Computation of entropy via IFS}

In this section we present a method of computing the dynamical entropy
(\ref{hdyn}), which is especially useful when the number of cells $k$ of the
partition of the space $X$  is large. We use the concept of iterated
function systems (IFSs), discussed in details in the book of Barnsley
\cite{Ba93}. Consider the set of $k$ functions $p_i: Y \mapsto{\mathbb R}^+$
and maps $F_i: Y \mapsto Y$ defined as \cite{S97,S99}
\begin{equation}
\left\{\begin{array}{rcl}
p_i(z)&=&\tau^TD(i)z \\
F_i(z)&=&\frac{D(i)z}{p_i(z)}\end{array}\right.\ \ i=1,\ldots,k\,,z\in Y\,,
\label{ifs1}
\end{equation}
where the vector $\tau$ and the matrices $D(i)$ are defined in Sects.~IIIB
and IVA, respectively, and $Y\subset{\mathbb R}^{N+1}$ is a convex closure
of the set of all vectors of the form $u(f(x))$ for $x\in[0,1]$.

Let us stress that the spaces $X=[0,1]$ and the $N+1$ dimensional space $Y$
are different. The normalization of the kernel $\int_X {\cal P}(f(x),y) dy
=1$ for $x\in X$, leads to the condition $\sum_{i=1}^k p_i(z)=1$ for any
$z\in Y$. Therefore the functions $p_i$ can be interpreted as place
dependent probabilities and together with the functions $F_i$ form an IFS.
It is uniquely determined by the dynamical system $f$ with the noise given
by the density ${\cal P}$ and a specific $k$--elements partition ${\cal A}$.
Thus, the number of cells $k$ determines the size of IFS. It can be shown
\cite{S99} that the entropy of the considered dynamical system with noise is
equal to the entropy of the associated IFS.

The IFS generates a Markov operator ${\cal M}$ acting on the space of all
probability measures on $Y$. For any measurable set $S \subset Y$ the
following equality holds
\begin{equation}
({\cal M}\nu)(S)=\sum_{i=1}^k\int_{F_i^{-1}(S)}p_i(w)d\nu(w).
\end{equation}
It describes the evolution of the measure $\nu$ transformed by ${\cal M}$.
If the functions $F_i$ fulfill the strong contraction conditions
\cite{Ba93}, there exists a unique attracting invariant measure $\mu$ for
this IFS
\begin{equation}
{\cal M}\mu=\mu\,,
\end{equation}
which, in general, displays multifractal properties \cite{SKZ98}. The total
entropy can be computed as the Shannon entropy
$h_k(p_1,\ldots,p_k)=-\sum_{i=1}^k p_i\ln p_i$ averaged over the invariant
measure \cite{KSZ97,S97}
\begin{equation}
H_{tot}(f_{\sigma};{\cal A}) = \int_Y h_k(p_1(y),\ldots, p_k(y))d\mu(y)\,.
\end{equation}
The calculation of such an integral from the definition corresponds to the
matrix method presented previously. However, the existence of the attracting
invariant measure $\mu$ and the Kaijser--Elton ergodic theorem
\cite{Ka81,El87} assures that
\begin{equation}
H_{tot}(f_{\sigma};{\cal A}) =
\lim_{n\rightarrow\infty}\frac{1}{n}\sum_{l=0}^{n-1}h_k(y_l),
\label{keiser}
\end{equation}
where $\{y_{l}\}$ is a generic random sequence produced by the IFS. Such a
method of computing of an integral is often called random iterated algorithm
\cite{Ba93}. We start computations from an arbitrary initial point $y_0$,
iterate it by the IFS, and compute the average (\ref{keiser}) along a random
trajectory. To avoid transient dependence on the initial point $y_0$ one
should not take into account a certain number of initial iterations. Note
that the computing time grows only linearly with the number of cells $k$ and
one does not need to perform the burdensome time limit (\ref{padep}).

We used a similar method to compute the quantum coherent states entropy
\cite{KSZ97} and the R\'{e}nyi entropies for certain classical deterministic
maps \cite{SKZ98}.

\section{Dynamical entropy for noisy systems - exemplary results}

In this section we will study the entropy of the R\'{e}nyi map and the
logistic map perturbed by the trigonometric noise given by
(\ref{trignoise}). We will consider the partitions ${\cal A}_k$ of the
interval $[0,1]$ into $k$ equal subintervals. We put $H(k):=H({\cal A}_k)$,
$H_{tot}(N,k):= H_{tot}(f_N;{\cal A}_k)$, $H_{noise}(N,k):=
H_{noise}(f_N;{\cal A}_k)$, $H_{dyn}(N,k):= H_{dyn}(f_N;{\cal A}_k)$ and
$H_{dyn}(N):=H_{dyn}(f_N)$.

\subsection{Boltzmann--Gibbs entropy}

A simple integration allows us to obtain the BG--entropy $H_{BG}(N)$ for
this kind of  the noise
\begin{equation}
H_{BG}(N)=-\int_0^1 \! d\xi \, {\rm C_N}\cos^N(\pi \xi)\ln\bigl[
{\rm C_N}\cos^N(\pi\xi) \bigr] =
{N\over 2} \left[\Psi(\textstyle\frac{N}{2}\displaystyle)-
\Psi(\textstyle\frac{N+1}{2}\displaystyle)\right]+1-\ln {\rm C_N},
\label{bgtrig}
\end{equation}
where $\Psi$ denotes the digamma function \cite{GR65} and the normalization
constant ${\rm C_N}$ is given by (\ref{cnorm}).

It follows from (\ref{bgtrig}) that in the deterministic limit
($N\rightarrow\infty$) the BG--entropy diverges to minus infinity, namely
\begin{equation}
\lim_{N\rightarrow\infty}-\frac{H_{BG}(N)}{\ln N} = \frac{1}{2}\,.
\end{equation}
This relation shows, how the maximal dynamical entropy $-H_{BG}$ (see
(\ref{bgineq2})), admissible by a certain level of the noise, grows
logarithmically in the deterministic limit.

\subsection{Entropy of the noise}

\begin{figure}
\centering
\setlength{\epsfxsize}{0.6\textwidth}
\epsfbox{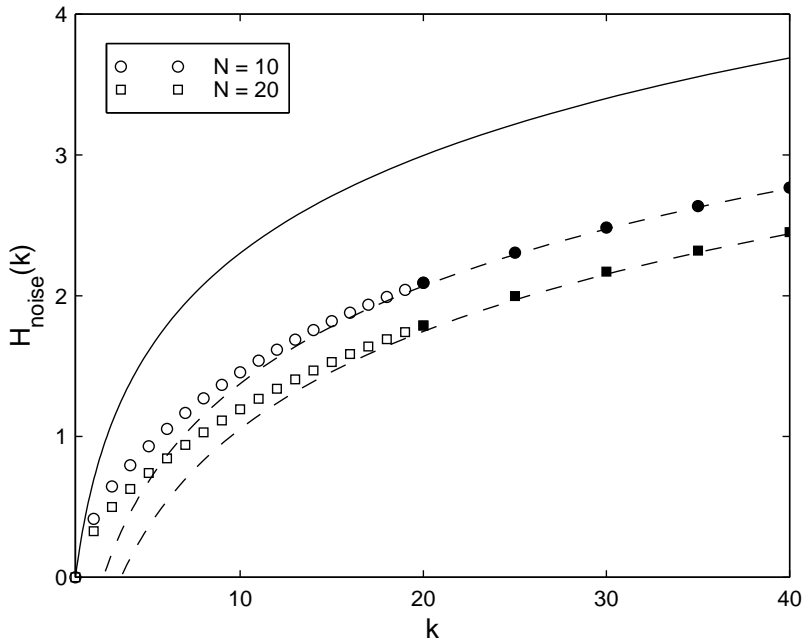}
\caption{Dependence of the entropy of the noise $H_{noise}(k)$ on the number
of cells $k$ in partition for $N=10 (\circ)~\mbox{and}~20(\Box)$.  Open
symbols are obtained with the matrix method, while the data for $k \geq 20$
are received with the IFS technique. Solid line represents the upper bound
($H_{{\cal A}_k}$) while two dashed lines provide lower bounds given by
(\protect{\ref{bgineqnoise}}).}
\label{hszifs}
\end{figure}

We used the matrix method of computing probabilities, which lead to partial
entropies $H_n$ and the relative entropies $G_n$. Fast (presumably
exponential) convergence of the sequence $G_n$ allowed us to approximate the
entropy by $G_7$ with accuracy of order $\approx10^{-6}$. Fig.~\ref{hszifs}
presents the dependence of the entropy of the noise $H_{noise}(N,k)$ on the
number of cells $k$ in the partition ${\cal A}_k$ for two different
amplitudes of noise ($N=10$ and $20$). The data for large number of cells
$(k \ge 20)$ are obtained by the technique of IFS. The results are compared
with the upper and lower bounds for the entropy of the noise which occurred
in (\ref{bgineqnoise}). It follows from (\ref{bgineqnoise}) that the entropy
diverges logarithmically with the number of cells $k$ in the partition. For
a fixed partition it decreases to zero with decreasing strength of the
stochastic perturbation (increasing parameter $N$).
%FIG 6

\subsection{Entropy for the noisy R\'{e}nyi map}

\begin{figure}
\centering
\setlength{\epsfxsize}{0.7\textwidth}
\epsfbox{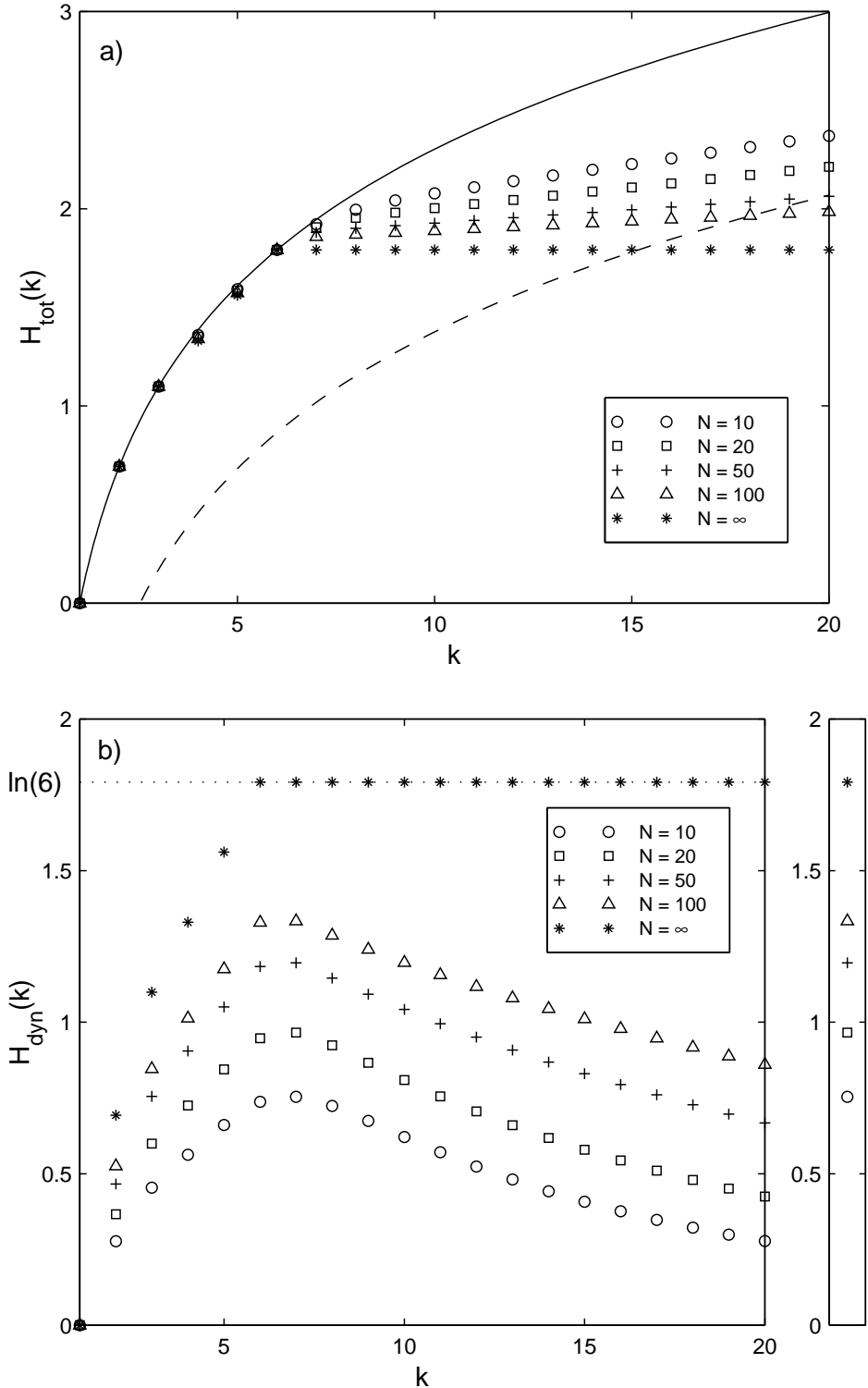}
\caption{Entropies for the R\'{e}nyi map $f(x)=[6 x]_{\rm mod~1}$ perturbed
by the noise with $N=10 (\circ), 20(\Box), 50(+)$ and $100 (\triangle)$ as a
function of the number of cells $k$: a) The total entropy $H_{tot}(k)$.
Solid line represents the upper bound ($H_{{\cal A}_k}$) while the dashed
line provides the lower bound (\protect{\ref{bgineq}}) for $N=10$. b) The
dynamical entropy $H_{dyn}(k)$. The maximum of each curve gives $H_{dyn}$ as
represented on the right side.}
\label{6xtot}
\end{figure}

The R\'{e}nyi map  $f_{(s)}(x)=[sx]_{\rm mod 1}$ ($s \in \mathbb N$), with
explicitly known metric entropy $H_{KS}(f_{(s)})=\ln{s}$, is particularly
suitable to test changes of the dynamical entropy with stochastic
perturbation. Results obtained for the trigonometric noise (\ref{trignoise})
are much more accurate than these obtained for rectangular noise and
reviewed briefly in Sec.~IID. Data presented below are received for the
R\'{e}nyi map with $s=6$ (we put $f=f_{(6)}$). Dependence of the total
entropy $H_{tot}(N,k)$ on the number of cells $k$ is presented in
Fig.~\ref{6xtot}a for four levels of noise ($N=10,20,50$ and $100$). The
solid line represents the entropy of the partition $H(k) = ln{k}$ (upper
bound) and the dashed line provides the $N$--dependent lower bound given by
$H(k) + H_{BG}(N)$ (for $N=10$), while the stars denote the partition
dependent entropy of the deterministic system given by $f$. It saturates at
the generating partition $k_g=6$ and achieves the value $H_{KS}(f) = ln(6)$.
It seems that this value gives an another lower bound for the total entropy
$H_{tot}$.

The total entropy and the entropy of the noise diverge in the limit of fine
partition ${\cal A} \downarrow 0$ ($k \rightarrow \infty$), but their
difference remains bounded.
%FIG 7
Fig.~\ref{6xtot}b shows the difference $H_{dyn}(N,k) = H_{tot}(N,k)-
H_{noise}(N,k)$ necessary for computation the dynamical entropy
(\ref{hdynsup}). This quantity tends to zero for $k=1$ and $k \to \infty$
(\ref{limdyn}) and achieves its maximum - giving a lower bound for the
dynamical entropy $H_{dyn}(N)$ - close to the number of cells $k_g$ in the
generating partition. Dynamical entropy is equal to zero for $N=0$ and
increases with the decreasing noise strength. In the limit $N\to \infty$ it
is conjectured to tend to the KS-entropy of the deterministic system
$H_{KS}= \ln(6)$ represented by a horizontal line.

\subsection{Entropy for the noisy logistic map}

\begin{figure}
\centering
\setlength{\epsfxsize}{0.6\textwidth}
\epsfbox{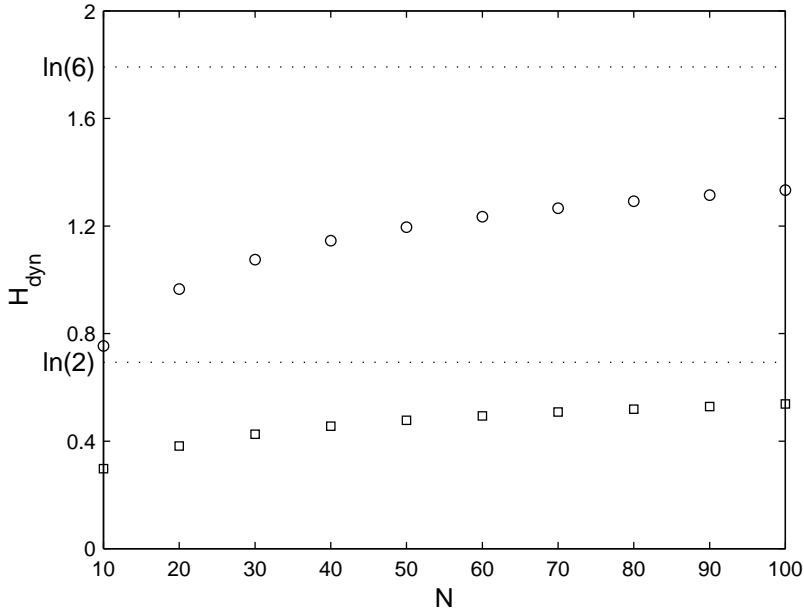}
\caption{Dynamical entropy $H_{dyn}$ for the R\'{e}nyi map $(\circ)$ and the
logistic map $(\Box)$ depicted as functions of the noise parameter $N$.
Horizontal lines represent the values ln(6) and ln(2) of the KS-entropy for
the corresponding deterministic maps.}
\label{hdynlim}
\end{figure}

A similar study was performed for the logistic map given by $f(x)=4x(1-x)$
for $x \in [0,1]$ perturbed by the trigonometric noise (\ref{trignoise}). As
before we treat the interval $X$ as a circle setting $f(x)=f(x_{\bmod 1})$. Numerical data produce pictures analogous to those obtained for the
R\'{e}nyi map with $s=6$. Instead of presenting them here, we supply a
compilation of the results for both systems. Computing total entropy and
entropy of the noise for several partitions we took the largest difference
between them as an approximation of the dynamical entropy (\ref{hdynsup}).
Fig.~\ref{hdynlim} shows how the dynamical entropy changes with the noise
parameter $N$ for both systems. It is conjectured to tend to the
corresponding values of the KS-entropy (ln(2) for the logistic map and ln(6)
for the R\'{e}nyi map) in the deterministic limit $N \to \infty$.
%FIG 8

\section{Concluding remarks}

The standard definition of the Kolmogorov--Sinai entropy is not applicable
for systems in the presence of a continuous random noise, since the
partition dependent entropy diverges in the limit of a fine partition. We
generalize the notion of the KS-entropy for dynamical systems perturbed by
an uncorrelated additive noise. The total entropy of a random system is
split into two parts: the {\sl dynamical entropy} and the {\sl entropy of
the noise}. In the deterministic limit (the variance of the noise tends to
zero) the entropy of the noise vanishes, while the dynamical entropy of the
stochastically perturbed system is conjectured to tend to the KS-entropy of
the deterministic system.

The continuous Boltzmann-Gibbs entropy characterizes the density of the
distribution of the noise. It provides an upper bound for the maximal
dynamical entropy observable under the presence of this noise. If the
BG--entropy is equal to zero such a noise may be called disruptive, because
one cannot draw out any information concerning the underlying deterministic
dynamics. Investigating properties of the dynamical entropy we find that the
two limits: the diameter of the partition to zero and the noise strength to
zero do not commute, and point out some consequences of this fact.

Computation of the dynamical entropy becomes easier, if one assumes that the
density of the noise can be expanded in a finite basis consisting of
continuous base functions. In this case we find a simple way of computing
the probabilities of trajectories passing through a given sequence of the
cells in the partition. The calculations are based on multiplication of
matrices of size $N+1$ and the computing time grows linearly with the length
of a trajectory $n$. On the other hand, diminishing the noise strength
causes an increase of the matrix dimension.

For each dynamical system perturbed by this kind of noise and for a given
$k$-element partition of the phase space we construct an associated iterated
function system, which consists of $k$ functions with place dependent
probabilities and acts in a certain $N+1$ dimensional auxiliary space.
Entropy of the dynamical system with noise is shown to be equal to the
entropy of IFS, which can be easily computed by the random iterated
algorithm. This method is particularly suitable for large number of cells
$k$, for which the number of possible trajectories grows in time as $k^n$.

We study some one--dimensional maps perturbed by trigonometric noise, for
which the basis functions are given by trigonometric functions. In this case
we can represent the Frobenius-Perron operator for the noisy system by a
matrix of size $N+1$. Diagonalizing this matrix numerically we find the
spectrum of this operator. Analysing the logistic map subjected by such a
random perturbation we indicate that the invariant measure tends to the
invariant measure of the deterministic system in the limit $N\to\infty$. On
the other hand, the spectrum of the Frobenius--Perron operator describing
the noisy system need not to tend to the corresponding characteristics of
the deterministic system.

The deterministic limit $N\to\infty$ resembles in a sense the semiclassical
limit of quantum mechanics $\hbar \to 0$. For example, if one discuss the
quantum analogues of classical maps on the sphere \cite{Ha91}, the size of
the Hilbert space $2j+1$ behaves as $1/\hbar$, where $j$ is the spin quantum
number. Therefore, it would be interesting to analyze such two--dimensional
classical systems in the presence of noise (in the case of two--dimensional
``trigonometric'' noise the FP-operator can be represented by a matrix of
the size $N^2$) and to compare, how the spectrum of a given classical map is
approached in two complementary limits: the semiclassical limit $j\to
\infty$ of the corresponding quantum map and the deterministic limit $N\to
\infty$ of a classical noisy system. Some preliminary results on related
issue of truncating the infinite matrix which represents the FP-operator of
a deterministic system have been achieved very recently \cite{Ha,Fi}.

One of us (K{\.Z}) would like to thank Ed Ott for the hospitality at the
University of Maryland, where a part of this work has been done and
gratefully acknowledges the Fulbright Fellowship. This work was also
supported by the Polish KBN grant P03B 060 13.

\appendix
\section{Exemplary Iterated Function System}
\label{matrixmet}

To illustrate the IFS method we discuss the computation of the entropy of
the noise given by (\ref{trignoise}) for $N=2$ and for the partition of the
interval $[0,1]$ into $k=4$ equal cells. In this case (\ref{ifs1}) gives the
IFS consisting of $k=4$ functions acting in a $3$-dimensional space $Y
\subset [-1,1]^3$. The probabilities $p_i$ are place dependent
\begin{equation}
\begin{array}{rcl}
p_1(x,y,z)=\frac{x(\pi+2)}{16\pi}+\frac{y}{8\pi}+
\frac{z(\pi-2)}{16\pi}&&
p_2(x,y,z)=\frac{x(\pi-2)}{16\pi}+\frac{y}{8\pi}+
\frac{z(\pi+2)}{16\pi}\\
p_3(x,y,z)=\frac{x(\pi-2)}{16\pi}-\frac{y}{8\pi}+
\frac{z(\pi+2)}{16\pi}&&
p_4(x,y,z)=\frac{x(\pi+2)}{16\pi}-\frac{y}{8\pi}+
\frac{z(\pi-2)}{16\pi} , \end{array}
\end{equation}
while the functions read
% &\hspace*{-9pt}
\begin{equation}
\begin{array}{l}
F_1(x,y,z)=\frac{1}{p_1(x,y,z)}
[\frac{x(8+3\pi)}{32\pi}+\frac{3y}{16\pi}+\frac{z}{32}\,;
\frac{3x}{8\pi}+\frac{y}{16}+\frac{z}{8\pi}\,;
\frac{x}{32}+\frac{y}{16\pi}+\frac{z(3\pi-8)}{32\pi}]\\
F_2(x,y,z)=\frac{1}{p_2(x,y,z)}
[\frac{x(3\pi-8)}{32\pi}+\frac{y}{16\pi}+\frac{z}{32}\,;
\frac{x}{8\pi}+\frac{y}{16}+\frac{3z}{8\pi}\,;
\frac{x}{32}+\frac{3y}{16\pi}+\frac{z(3\pi+8)}{32\pi}]\\
F_3(x,y,z)=\frac{1}{p_3(x,y,z)}
[\frac{x(3\pi-8)}{32\pi}-\frac{y}{16\pi}+\frac{z}{32}\,;
-\frac{x}{8\pi}+\frac{y}{16}-\frac{3z}{8\pi}\,;
\frac{x}{32}-\frac{3y}{16\pi}+\frac{z(3\pi+8)}{32\pi}]\\
F_4(x,y,z)=\frac{1}{p_4(x,y,z)}
[\frac{x(3\pi+8)}{32\pi}-\frac{3y}{16\pi}+\frac{z}{32}\,;
-\frac{3x}{8\pi}+\frac{y}{16}-\frac{z}{8\pi}\,;
\frac{x}{32}-\frac{y}{16\pi}+\frac{z(3\pi-8)}{32\pi}]  \\
\end{array}
\end{equation}
for $(x,y,z) \in Y$.

\begin{figure}
\centering
\setlength{\epsfxsize}{0.6\textwidth}
\epsfbox{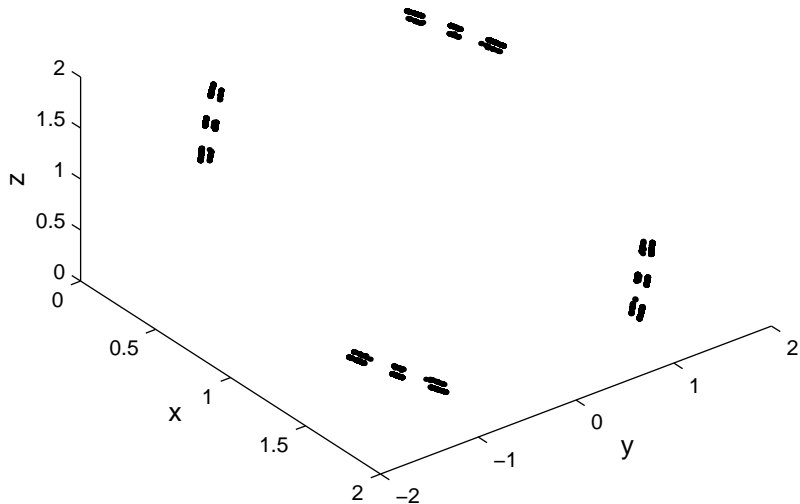}
\caption{Fractal support of the invariant measure $\nu_{\star}$ of the IFS
associated with the trivial dynamical system $f(x)=x$ in the presence of the
trigonometric noise with $N=2$. The number of cells $k=4$ determines the
number of functions in the IFS and the structure of the depicted set.}
\label{ifsdense}
\end{figure}

Fig.~\ref{ifsdense} presents the support of the invariant measure $\mu$ for
this IFS.  Applying the random iteration algorithm we obtain in this case
the entropy of the noise $H_{noise}=1.1934$. Ironically, less interesting
(more contracting) fractal picture leads to a faster convergence of the sum
(\ref{keiser}).
%FIG 9

\end{document}